\documentclass{aa}

\usepackage[varg]{txfonts}
\usepackage{siunitx}
\usepackage[version=4]{mhchem}
\usepackage{multirow}
\usepackage{graphicx}
\usepackage[caption=false]{subfig}
\usepackage[strict]{changepage}
\usepackage{xcolor}
\usepackage{ulem}
\usepackage[colorlinks=true,citecolor=blue]{hyperref}
\usepackage{float}
\usepackage{amssymb}
\usepackage{amsmath}
\usepackage{lscape}
\usepackage{mathtools}
\usepackage{natbib}
\usepackage{pdflscape}
\usepackage{gensymb}

\defcitealias{zhou2023}{Paper~I}   

\DeclareSIUnit\jansky{Jy}

\newcolumntype{L}[1]{>{\raggedright\let\newline\\\arraybackslash\hspace{0pt}}m{#1}}
\newcolumntype{C}[1]{>{\centering\let\newline\\\arraybackslash\hspace{0pt}}m{#1}}
\newcolumntype{R}[1]{>{\raggedleft\let\newline\\\arraybackslash\hspace{0pt}}m{#1}}

\begin{document}

\title{Physical properties of embedded clusters in ATLASGAL clumps with HII regions}
\author{J. W. Zhou\inst{\ref{inst1}} 
\and Pavel Kroupa\inst{\ref{inst2}}
\fnmsep \inst{\ref{inst3}}
\and Sami Dib \inst{\ref{inst4}}
}
\institute{
Max-Planck-Institut f\"{u}r Radioastronomie, Auf dem H\"{u}gel 69, 53121 Bonn, Germany \label{inst1} \\
\email{jwzhou@mpifr-bonn.mpg.de}
\and
Helmholtz-Institut f{\"u}r Strahlen- und Kernphysik (HISKP), Universität Bonn, Nussallee 14–16, 53115 Bonn, Germany \label{inst2}\\
\email{pkroupa@uni-bonn.de}
\and
Charles University in Prague, Faculty of Mathematics and Physics, Astronomical Institute, V Hole{\v s}ovi{\v c}k{\'a}ch 2, CZ-180 00 Praha 8, Czech Republic \label{inst3}
\and
Max Planck Institute f\"{u}r Astronomie, K\"{o}nigstuhl 17, 69117 Heidelberg, Germany \label{inst4}\\
\email{dib@mpia.de}
}

\date{Accepted XXX. Received YYY; in original form ZZZ}

\abstract
{
Using the optimal sampling model, we synthesized the embedded clusters of ATLASGAL clumps with HII regions (HII-clumps). The 0.1 Myr isochrone was used to estimate the bolometric luminosity of each star in an embedded cluster, we also added the accretion luminosity of each star in the embeded cluster. The total bolometric luminosity of synthetic embedded clusters can well fit the observed bolometric luminosity of HII-clumps.
More realistically, we considered the age spread in the young star and protostar populations in embedded clusters of HII-clumps by modeling both constant and time-varying star formation histories (SFHs). According to the age distribution of the stellar population, we distributed the appropriate isochrones to each star, and sorted out the fraction of stellar objects that are still protostars (Class 0 and Class I phases) to properly add their accretion luminosities.
Compared to a constant SFH, burst-like and time-dependent SFHs can better fit the observational data. We found that as long as 20\% of the stars within the embedded cluster are still accreting, the contribution of accretion luminosity will be significant to the total bolometric luminosity of low-mass HII-clumps with mass log$_{10}$(M$_{\rm cl}$/M$_{\odot}$) $<$ 2.8. 
Variations in the accretion rate, the SFE and the initial mass function (IMF) and more physical processes like the external heating from HII regions and the flaring from pre-main sequence (PMS) stars and protostars need to be investigated to further explain the excess luminosity of low-mass HII-clumps.
}


\titlerunning{}
\authorrunning{J. W. Zhou, Pavel Kroupa, Sami Dib}

\maketitle 

\section{Introduction}\label{intro}

Until a decade ago, the fields of stellar dynamics and chemical abundances on the one hand and star cluster formation on the other were operating almost independently with the stellar kinematics and abundances community focusing on the long-term dynamical evolution of stellar clusters \citep{aarseth1974-35, heggie2006-368,Lamers2008-674,PortegiesZwart2010-48,Spurzem2018-990}, while the star-formation community has directed its attention toward the dynamics of interstellar gas and the impacts of star formation and stellar feedback \citep{bate1998-508, dib2008-678, smith2009-400, dale2017-467, Motte2018-56, zhou2024-682}. As highlighted in \citet{Krumholz2019-57}, maintaining this division is no longer tenable. Advances in observational capabilities and large surveys of the populations of young and less young clusters in the Milky Way now allow us to trace the origins of field stars back to their birthplaces, or at least reconstruct some of the now-dissolved structures in which they originated \citep{wright2023-arxiv, cantatgaudin2024-683}. To guide such reconstructions, input from both theoretical and observational perspectives within the star-formation community is essential.

However, our understanding of massive stars and star clusters formation remains incomplete in the present day. Most and perhaps all the observed stars were formed in embedded clusters \citep{Kroupa1995a-277, Kroupa1995b-277, Lada2003-41,Kroupa2005-576,Megeath2016-151, Dinnbier2022-660}, introducing considerable complexity in the identification and analysis of individual young stellar objects. Unlike their lower-mass counterparts, massive stars progress to the main sequence while still deeply embedded in their natal clumps, rendering all initial stages invisible even when observed at mid-infrared wavelengths (not all). Adding to the challenge, massive stars are both rare and undergo rapid evolution. Typically, the sites where they form are situated at much greater distances compared to their low-mass counterparts, further complicating observational efforts \citep{Urquhart2022-510, Wells2022-516, Reyes2024-529}.

Fortunately, our understanding of massive star and cluster formation has seen significant advancements, thanks to the combination of Galactic plane surveys and high-angular resolution images obtained through submillimeter facilities \citep{Motte2018-56}. Notably, surveys like HiGAL \citep{Molinari2010} and ATLASGAL \citep{Schuller2009-504} have played a crucial role by providing unbiased samples of dense clumps, identified by their dust emission. Dense clumps serve as the initial stages of massive star and cluster formation. Investigations utilizing extensive and statistically representative clump samples can offer valuable perspectives on the effectiveness of transforming molecular gas into stellar clusters. This approach enables us to evaluate how the physical environment influences this intricate process, providing a deeper understanding of the kinematics and dynamics involved \citep{Urquhart2022-510}.
In \citet{Urquhart2022-510}, 
5007 ATLASGAL clumps have been classified  into four evolutionary stages, i.e. quiescent, protostellar, young stellar objects (YSOs) and HII regions. In this work, we focus on the clumps at the most evolved stage, called HII-clumps. HII-clumps represent the final stage of embedded cluster formation in the clumps, which are critical to build the bridge between the stellar kinematics and abundances community and the star-formation community, because protoclusters or embedded clusters in HII-clumps are the first stage of gas-free star clusters. 
Currently, it is difficult to construct complete star catalogs for early-stage and distant star clusters from observations \citep{Feigelson2013-209,Romine2016-833,Kuhn2017-154}. 
If there is a continuous transition from protoclusters to embedded clusters \citep{Zhou2024arXiv240720150Z}, and assuming the stellar IMF is a universal optimal distribution function rather than a probability density \citep{Kroupa2013-115,Yan2023-670}\footnote{See however \citet{Marks2012-422,Dib2023-959} for how the slope of the IMF in the intermediate-to high-mass end can vary with the environment.}, 
we can directly infer the complete census of stars in HII-clumps. 
In this work we will investigate the physical properties of embedded clusters in HII-clumps and estimate the final star formation efficiency of the clumps.

\section{Samples}\label{sample}
\begin{figure}
\centering
\includegraphics[width=0.45\textwidth]{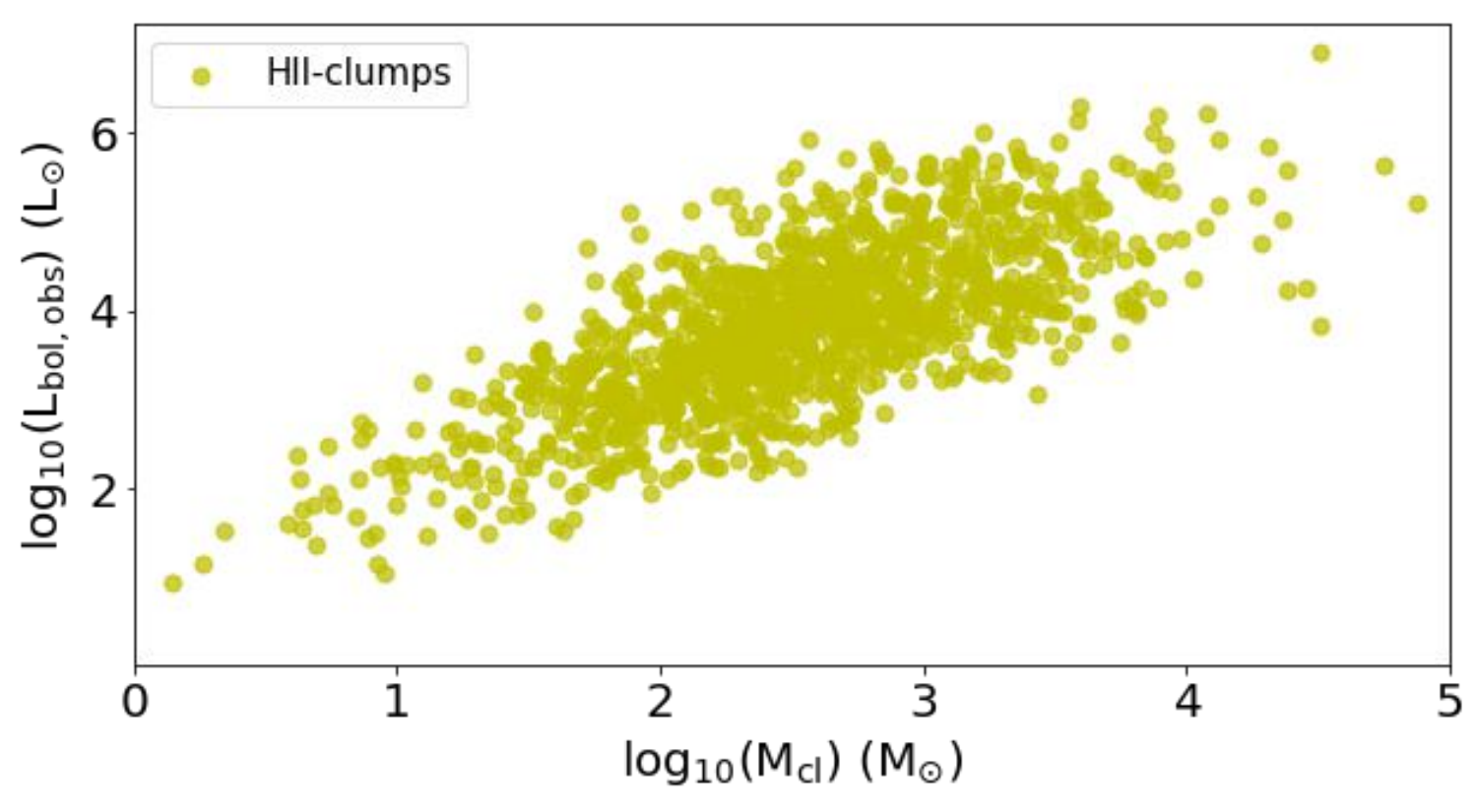}
\caption{The L$_{\rm bol, obs}$-M$_{\rm cl}$ relation of HII-clumps.}
\label{selection}
\end{figure}

The physical parameters of 1246 HII-clumps have been calculated and listed in Table.4 of \citet{Urquhart2022-510}. 
Clump sizes are determined based on the pixel count within the Full Width at Half Maximum (FWHM) contour, i.e. above 50 percent of the peak of the ATLASGAL dust continuum emission. The clump's mass, denoted as M$_{\rm cl}$ and measured at FWHM, is determined by integrating the flux density at 870 $\mu$m within the 50 percent contour, as outlined in \citet{Urquhart2022-510}.
The bolometric luminosity (L$_{\rm bol, obs}$) is derived from the spectral energy distributions (SEDs) of the clumps, which are constructed using continuum data spanning from mid-infrared to sub-millimeter wavelengths
\citep{Konig2017-599,Urquhart2022-510}.
There are 1226 HII-clumps with mass M$_{\rm cl}$ and bolometric luminosity L$_{\rm bol, obs}$ estimates, Fig.\ref{selection} shows their L$_{\rm bol, obs}$-M$_{\rm cl}$ relation.

\section{Results}
\subsection{Synthetic embedded clusters}\label{cluster}
\begin{figure}
\centering
\includegraphics[width=0.47\textwidth]{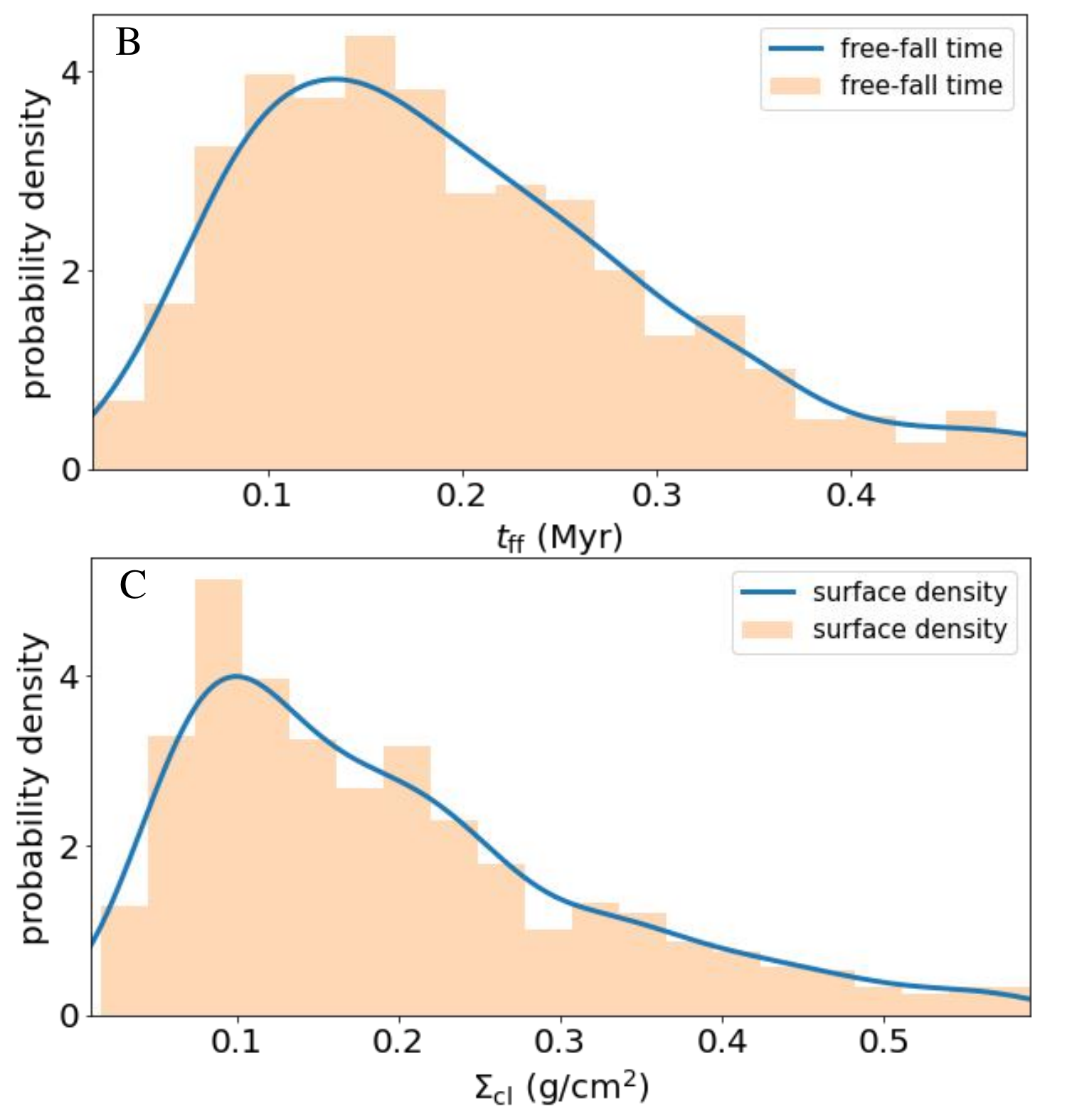}
\caption{Physical properties of HII-clumps. (a) Distribution of the free-fall time; (b) Distribution of the surface density.}
\label{quantity}
\end{figure}

Adopting the optimal sampling algorithm, we generate a stellar population for an embedded cluster with a mass denoted as M$_{\rm ecl}$ using the publicly accessible GalIMF code. A detailed description of the optimal sampling and of the GalIMF code that optimally populates clusters and galaxies with stars can be found in \citet{Schulz2015-582} and \citet{Yan2017-607}.
There are 3 input parameters to the code: the stellar mass M$_{\rm ecl}$ of the embedded cluster, the initial metallicity and the age of the embedded cluster. 
The metallicity we adopt here is the solar metallicity ([Fe/H] = 0). For the age,
\citet{Urquhart2022-510} calculated the free-fall times of all clumps, which provides a lower limit to the star formation time-scale. 
The free-fall time-scales of the clumps are a few times 10$^{5}$ yr, as shown in Fig.\ref{quantity}(a). 
Considering the embedded cluster in a HII-clump should be a mixture of pre-main-sequence (PMS) stars and protostars with spanning ages. The age in the code was taken to be 1 Myr.

The HII-clump mass M$_{\rm cl}$ and its embedded star cluster mass M$_{\rm ecl}$ satisfy M$_{\rm ecl}$ = SFE * M$_{\rm cl}$, where SFE is the final star formation efficiency of the clump.
In this work, we take SFE = 0.33. This value has been widely used in previous simulations and has been proven to be effective in reproducing the observational properties of stellar clusters \citep{Kroupa2001-321,
Banerjee2012-746,
Banerjee2013-764,
Banerjee2014-787,
Banerjee2015-447,
Oh2015-805,
Oh2016-590,
Banerjee2017-597,
Brinkmann2017-600,
Oh2018-481,
Wang2019-484,
Pavlik2019-626,
Wang2020-491,
Dinnbier2022-660}. 
Such a SFE is also consistent with the value obtained from
hydrodynamic calculations including self-regulation \citep{Machida2012-421,Bate2014-437} and as well with observations of embedded systems in the solar neighborhood \citep{Lada2003-41,Megeath2016-151,Zhou2024arXiv240720150Z}.
In Fig.\ref{selection}, the mass range is log$_{10}$(M$_{\rm cl}$/M$_{\odot}$) $\in$ [1, 5], and we take the values uniformly at a step size of 0.1 in this range. 
Then according to the corresponding M$_{\rm ecl}$, a population of stars is synthesized using the GalIMF code, which will be used to fit the data points in Fig.\ref{selection}. 

\subsection{Isochrone model}
\begin{figure}
\centering
\includegraphics[width=0.45\textwidth]{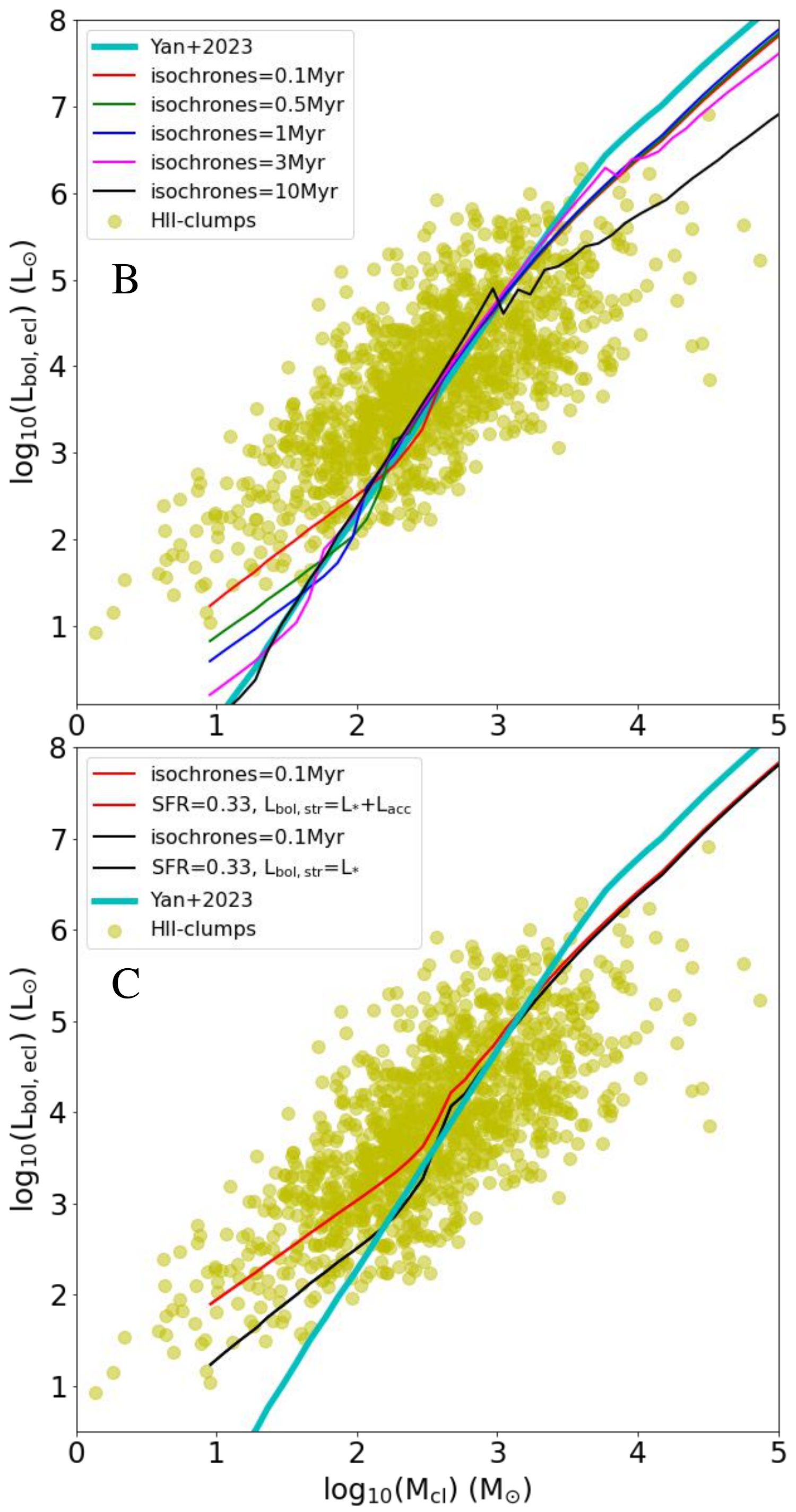}
\caption{Fitting the observed HII-clumps in \citet{Urquhart2022-510} using the optimal sampling model. (a) Assuming L$_{\rm bol, str}$=L$_{*}$, L$_{*}$ were estimated by different isochrones; (b) Assuming L$_{\rm bol, str}$=L$_{*}$+L$_{\rm acc}$, L$_{*}$ were estimated by the 0.1 Myr isochrone.}
\label{fit}
\end{figure}

\begin{figure}
\centering
\includegraphics[width=0.48\textwidth]{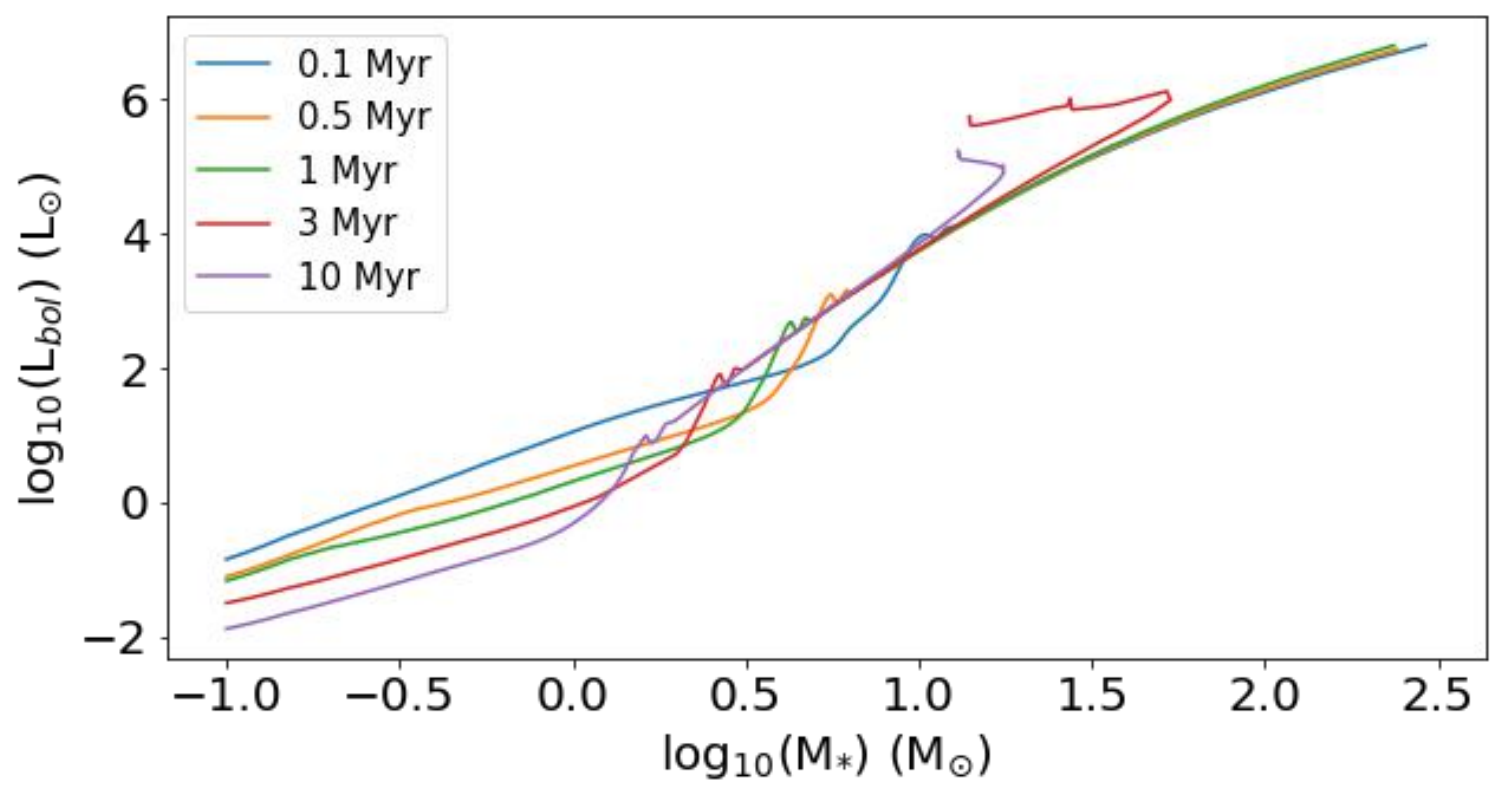}
\caption{The isochrones from the MESA Isochrones and Stellar Tracks (MIST) project.}
\label{isochrone}
\end{figure}

In \citet{Yan2023-670}, 
assuming that the pre-main sequence stars have a bolometric luminosity equal to the zero age main sequence (ZAMS) stars,
the bolometric luminosity of each optimally sampled star in a embedded star cluster was calculated by the empirical mass–luminosity relations summarised in Eq.(12) of \citet{Yan2023-670}. Then, the luminosity of all stars were added together to get the luminosity of the embedded cluster.
\citet{Yan2023-670} found that the optimal sampling model qualitatively agrees with the observation in \citet{Urquhart2022-510}. 
However, the observational data systematically become more luminous than the optimal sampling model for low-mass HII-clumps, as shown in Fig.\ref{fit}(a). 
Low-mass embedded clusters in low-mass HII-clumps are predominantly populated by low-mass stars. 
As these low-mass stars become observable within HII-clumps, they are still in the process of contracting onto the main sequence with significantly larger luminosities. Consequently, it is essential to incorporate their pre-main sequence evolutionary tracks into the optimal sampling model.
To achieve this goal,
we use pre-main sequence isochrones to estimate the bolometric luminosity for each star rather than using the empirical mass–luminosity relation. For a synthetic embedded cluster, the mass of each star is known. Using the isochrones that are displayed in Fig.\ref{isochrone}, 
the bolometric luminosity of a star can be estimated from its mass.

We use the isochrones from the MESA Isochrones and Stellar Tracks (MIST) project. The input physics, general overview of the models, and comparisons with observations are described extensively in \citet{Choi2016-823}.  
The MIST stellar evolutionary tracks are computed with the Modules for Experiments in Stellar Astrophysics (MESA) code \citep{Paxton2011-192,Paxton2013-208,Paxton2015-220}. The isochrone construction procedure is explained in detail in \citet{Dotter2016-222}.
The default set of ages in the isochrones ranges are from log$_{\rm 10}$ (Age/yr) = 5.0 to log$_{\rm 10}$ (Age/yr) = 10.3 with a step of 0.05.
Fig.\ref{isochrone} displays the isochrones of 0.1, 0.5, 1, 3 and 10 Myr.
As shown in Fig.\ref{fit}(a),
using the 0.1 Myr isochrone appears to best fit the observed data points. Although it is unlikely that all the stars in a HII-clump are that young.

\subsection{Accretion luminosity}\label{accretion}
\begin{figure}
\centering
\includegraphics[width=0.47\textwidth]{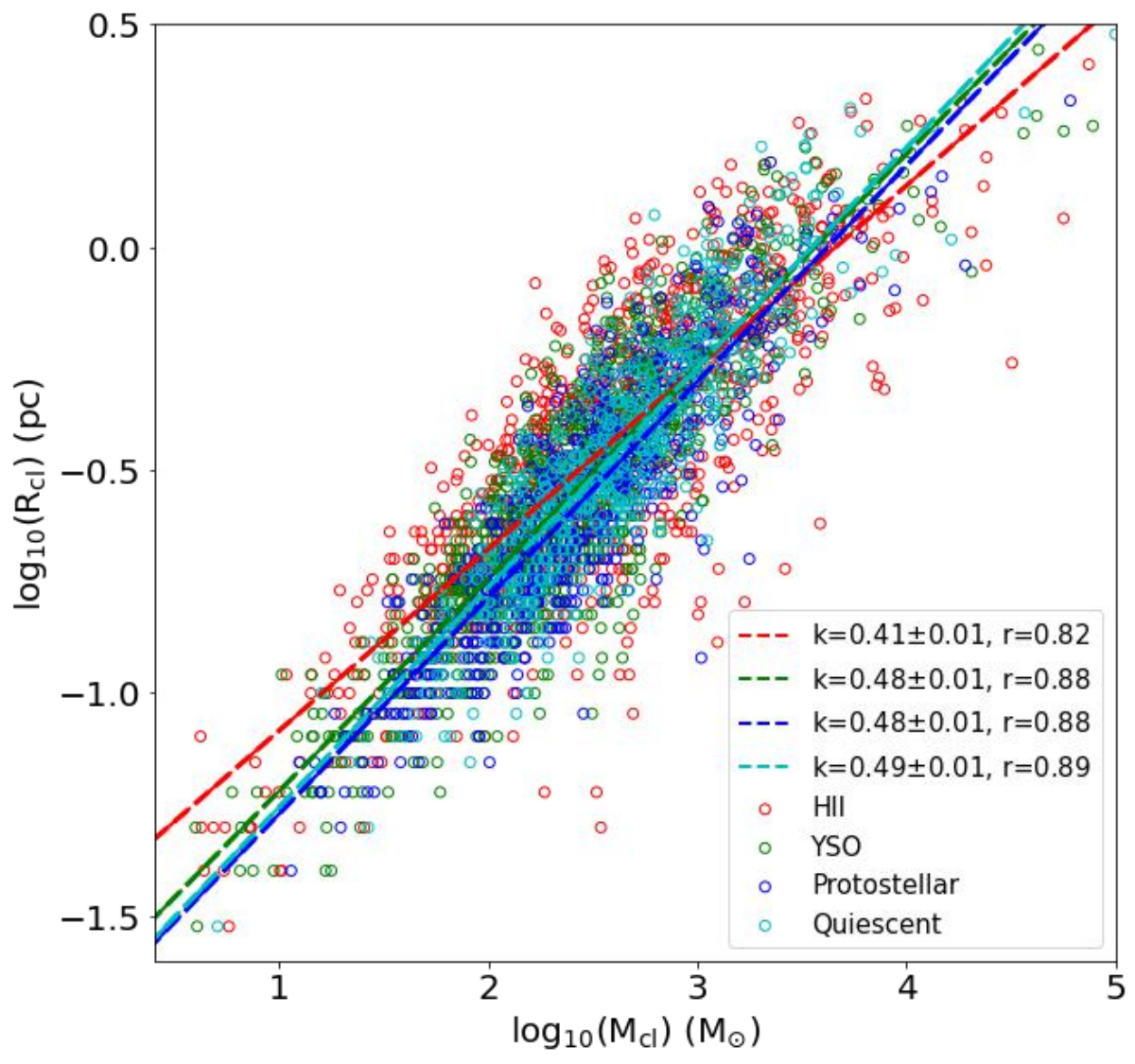}
\caption{The mass-size relations of the ATLASGAL clumps at four different evolutionary stages. ``k'' is the slope of the linear fitting.
``r'' represents the Pearson correlation coefficient.}
\label{lose}
\end{figure}

In Fig.\ref{fit}(a), after using the isochrones, the fitting to the data is better than that in \citet{Yan2023-670} who only considered the zero age main sequence luminosities of all stars in the embedded cluster. 
However, the models are still systematically lower than the observations for low-mass HII-clumps, even using the 0.1 Myr isochrone.
For HII-clumps, feedback from the HII region may cause part of the clump mass to be dispersed. Therefore, the observed data points will shift to the left, thus deviating from the models. In order to check this possibility, we compared the mass-size relations of the ATLASGAL clumps at four different evolutionary stages in Fig.\ref{lose}. 
The clumps in the three earlier evolutionary stages should not experience significant mass dispersal, maintaining a nearly identical mass-size relation. Being more evolved, HII clumps would exhibit a notable deviation from the initial mass-size relation if they had undergone significant mass dispersal. However, as shown in Fig.\ref{lose}, they only show a minor deviation, suggesting that they have not experienced significant mass dispersal.
Actually, as discussed in \citet{Urquhart2022-510}, this minor deviation may be attributed to observational biases.

Given that HII-clumps have ongoing star formation, the accretion luminosity L$_{\rm acc}$ from protostars may contribute to the total bolometric luminosity, and it is not explicitly included in the pre-main sequence evolutionary tracks. 
For each accreting protostar in a clump, its total bolometric luminosity L$_{\rm bol, str}$ is the sum of the accretion luminosity L$_{\rm acc}$ and the luminosity of the protostar L$_{*}$.
As shown in Fig.4 of \citet{Hosokawa2009-691}, when M$_{*}$<10 M$_\odot$, L$_{\rm acc}$ is much larger than L$_{*}$.

L$_{*}$ can be calculated using the isochrones. For the accretion, \citet{Davies2011-416} compared different accretion modes, and found that the prescription of accretion in \citet{McKee2003-585} excellently fits the observations. Thus, we compute L$_{\rm acc}$ using equations (41) and (73) in \citet{McKee2003-585}
\begin{equation}
L_{\rm acc}
= 4685 
\left(\frac{f_{\rm acc}}{0.5}\right) 
\left(\frac{M_{\rm *}}{30 M_{\odot}}\right) 
\left(\frac{\dot M_{\rm *}}{10^{-4} M_{\odot} yr^{-1}}\right)
\left(\frac{10 R_{\odot}}{r_{\rm *}}\right) L_{\odot}, 
\end{equation}
\begin{equation}
\dot M_{\rm *}=4.6 \times 10^{-4} \left(\frac{M_{\rm *f}}{30 M_{\odot}}\right)^{3/4}\Sigma_{\rm cl}^{3/4}\left(\frac{M_{\rm *}}{M_{\rm *f}}\right)^{0.5} M_{\odot} {\rm yr}^{-1},
\label{rate}
\end{equation}
where $f_{\rm acc}$ is the fraction of the protostar’s gravitational energy radiated by the star+disk, the fiducial value is 0.5. M$_{\rm *f}$ is the final mass of the star. M$_{\rm *}$ is at any time the mass of a star, $r_{\rm *}$ is its corresponding radius. $\Sigma_{\rm cl}$ is the surface density of the clump. 
The values of M$_{\rm *}$ and r$_{\rm *}$ can be estimated from the isochrones. From Fig.\ref{quantity}(b),
we take the surface density $\Sigma_{\rm cl} \approx$ 0.15 g cm$^{-2}$ as the fiducial value in our calculation.
We also make a simplified assumption, i.e. M$_{\rm *f} \approx$ M$_{\rm *}$.
As shown in Fig.\ref{fit}(b), after adding the accretion luminosity for each star, the model can fit the observed data.

\subsection{Age distribution}\label{age}


\begin{figure}
\centering
\includegraphics[width=0.45\textwidth]{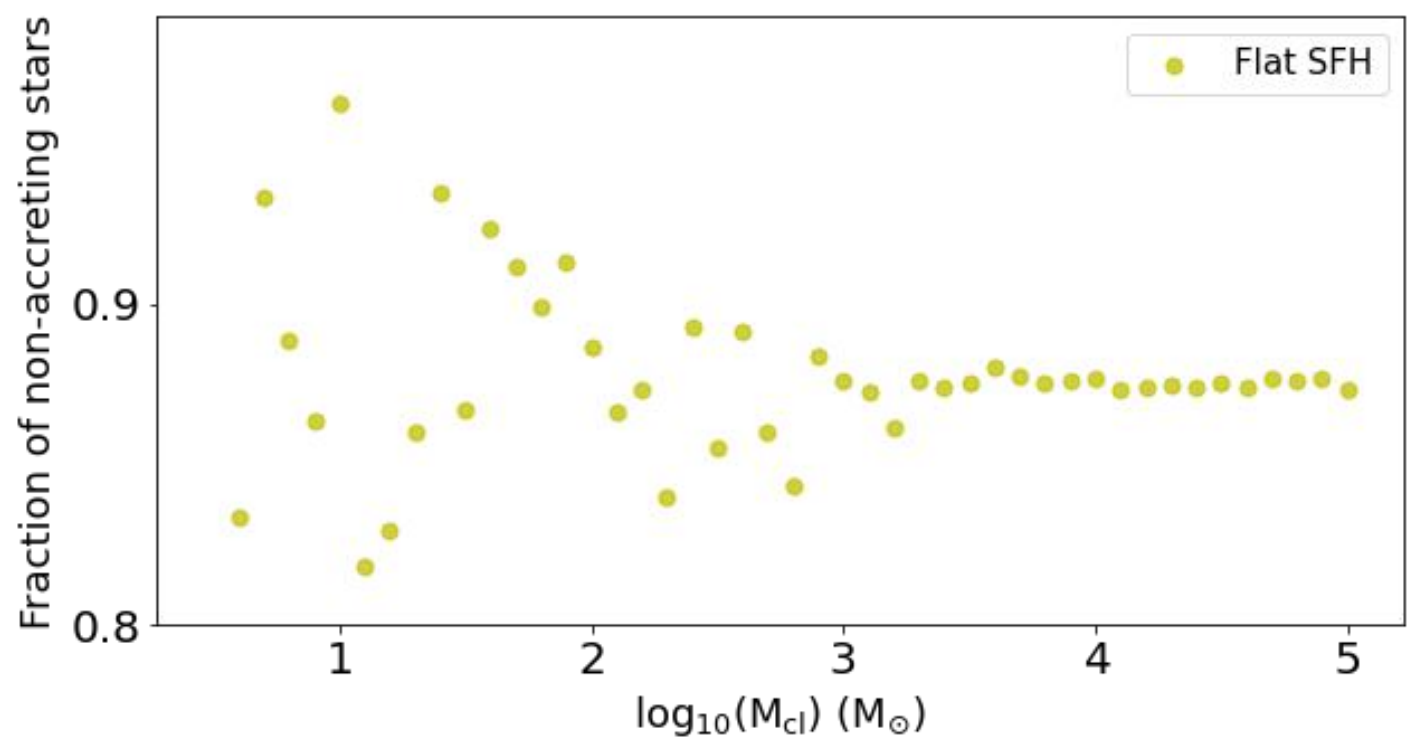}
\caption{Fraction of non-accreting stars in the case of a flat SFH.}
\label{flat}
\end{figure}

\begin{figure*}[htbp!]
\centering
\includegraphics[width=1\textwidth]{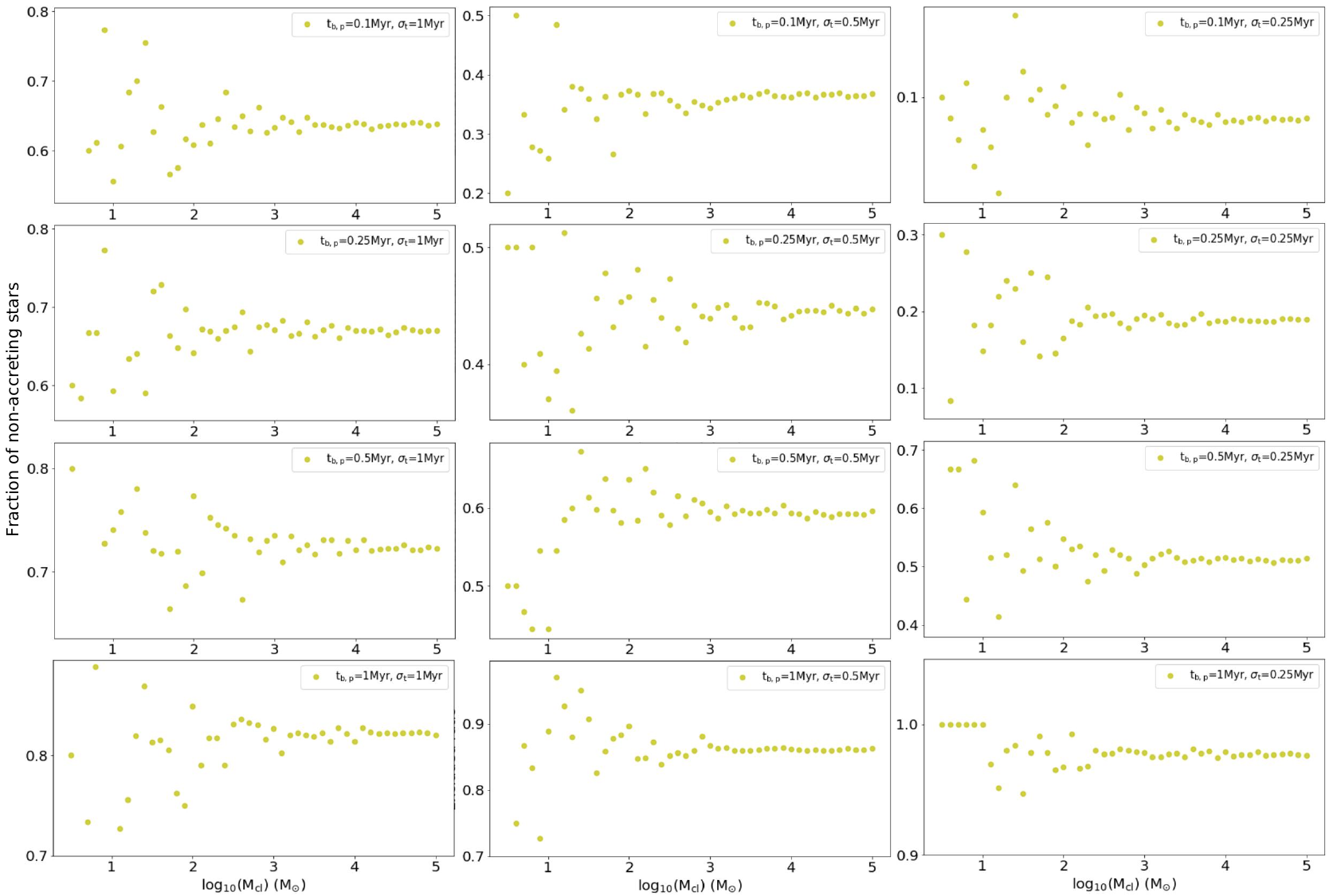}
\caption{Fraction of non-accreting stars in the case of a Gaussian SFH. From left to right, they are $\sigma_{\rm t}=$ 1, 0.5 and 0.25 Myr, respectively; From top to bottom, they are $t_{\rm b,p}$ = 0.1, 0.25, 0.5 and 1 Myr, respectively.}
\label{g1}
\end{figure*}

\begin{figure*}[htbp!]
\centering
\includegraphics[width=0.85\textwidth]{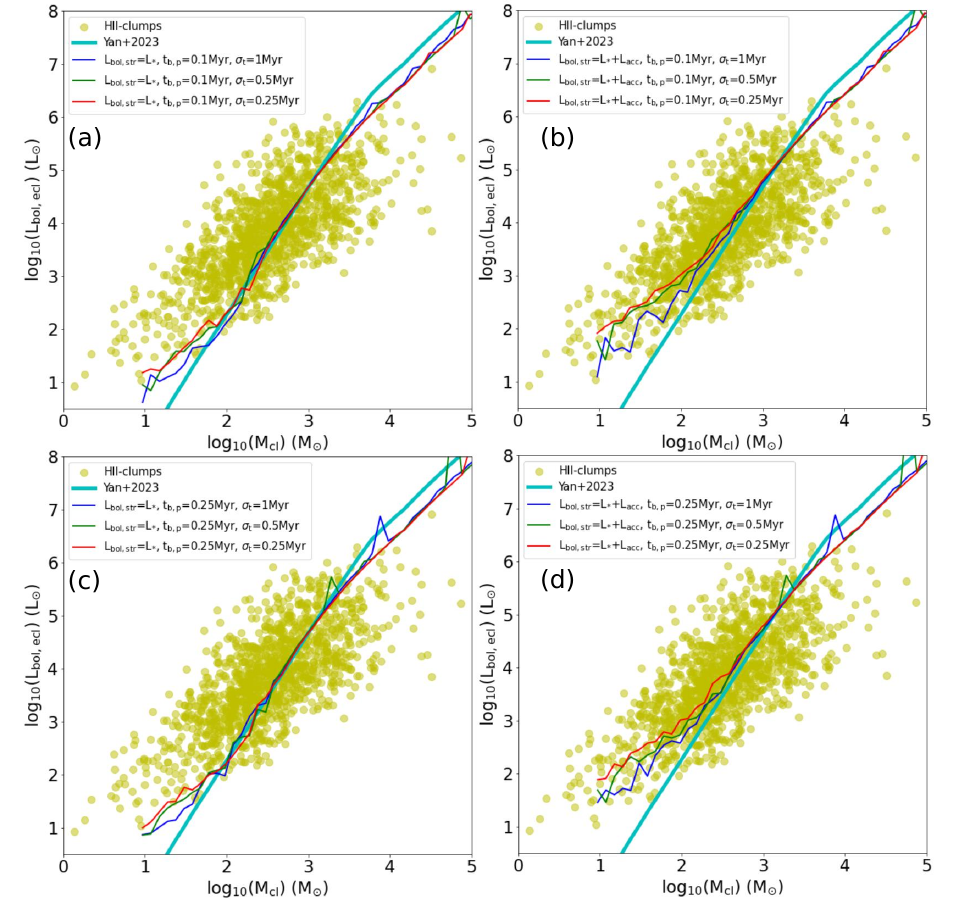}
\caption{Similar to Fig.\ref{fit}(b), but taking into account the burst-like and time-dependent star formation history (SFH) of the embedded clusters and the age distribution of their stellar populations. (a) $t_{\rm b,p}$ = 0.1 Myr and L$_{\rm bol, str}$=L$_{*}$; (b) $t_{\rm b,p}$ = 0.1 Myr and L$_{\rm bol, str}$=L$_{*}$+L$_{\rm acc}$; (c) $t_{\rm b,p}$ = 0.25 Myr and L$_{\rm bol, str}$=L$_{*}$; (d) $t_{\rm b,p}$ = 0.25 Myr and L$_{\rm bol, str}$=L$_{*}$+L$_{\rm acc}$.  L$_{\rm bol, str}$, L$_{*}$ and L$_{\rm acc}$ are the
total bolometric luminosity, the stellar luminosity and the accretion luminosity of the protostar, respectively.}
\label{g2}
\end{figure*}

\begin{figure*}[htbp!]
\centering
\includegraphics[width=0.85\textwidth]{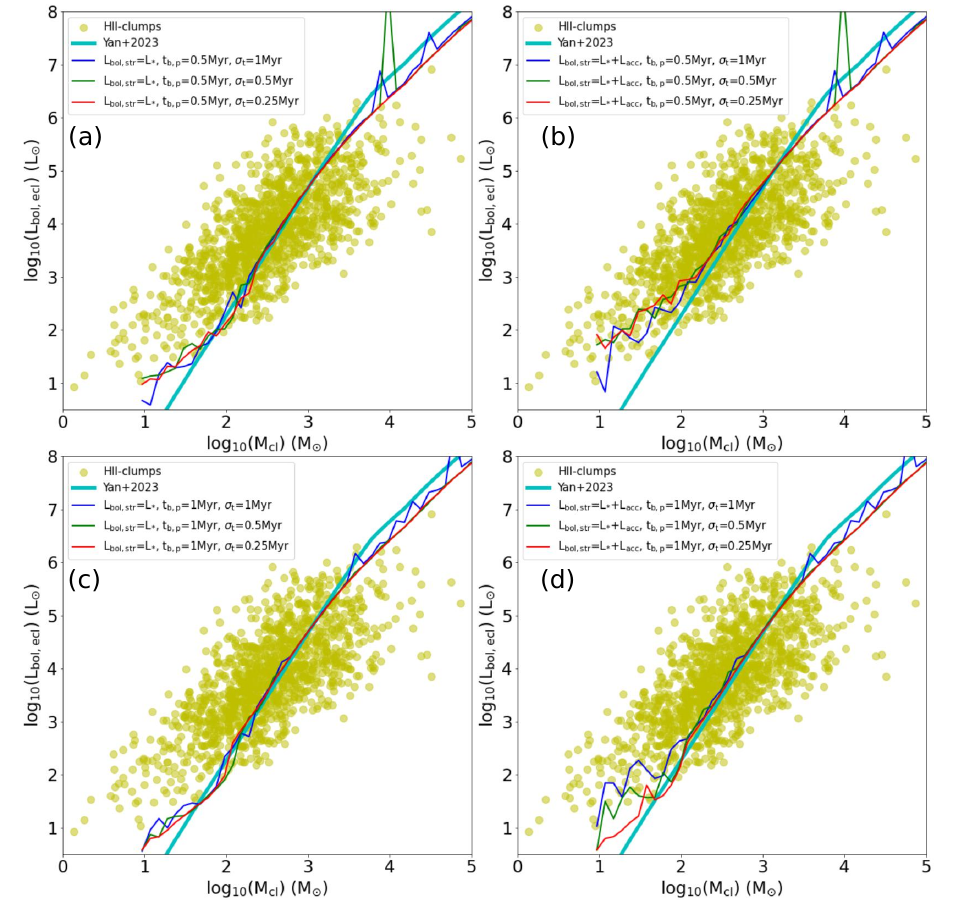}
\caption{Same as Fig.\ref{g2}, but for $t_{\rm b,p}$ = 0.5 Myr and $t_{\rm b,p}$ = 1 Myr.}
\label{g3}
\end{figure*}

In the above analysis, we only considered the 0.1 Myr isochrone. Actually, many stars in HII-clumps should be older than 0.1 Myr. 
There is certainly an age spread in the young star and protostar populations in embedded clusters of HII-clumps. Thus, we should choose the appropriate isochrones based on the ages of the stars. Moreover, not all stars in HII-clumps are in an accretion phase, which also depends on their ages. In order to address these issues, we use the method presented in \citet{Dib2024arXiv} to create an age distribution of the stellar population in each synthetic cluster to properly distribute the isochrone to each star. Then, we sorted out the fraction of stellar objects that are still protostars to properly add their accretion luminosities.

We examine scenarios where the star formation history (SFH) remains constant and others where it varies over time. In the cases of a constant SFH, we randomly sample the age of stars, $t_{\rm b}$, in the 0$-$4 Myrs time range. For a time-varying SFH, we employ a Gaussian function, allowing us to adjust both the peak position and the distribution width in the 0$-$4 Myrs time range. For this choice of the SFH, the star age, $t_{\rm b}$, is randomly drawn from the Gaussian distribution which is given by:
\begin{equation}
P(t_{\rm b})  = \frac{1} {\sigma_{\rm t} \sqrt{2 \pi}} \exp\left(-\frac{1}{2} \left(\frac{t_{\rm b}-t_{\rm b,p}}{\sigma_{t}}\right)^{2}\right),
\label{eq4}
\end{equation}
where $t_{\rm b,p}$ is the position of the peak and $\sigma_{\rm t}$ the standard deviation.

In this scenario of burst-like, time-dependent SFH, the $t_{\rm b}$ values of stars are randomly chosen based on the probability distribution function $P(t_{\rm b})$ as defined in equation.\ref{eq4}. 
In Fig.\ref{fit}(a), the 0.1 Myr isochrone can best fit the observed data points, indicating that most of the stars should be very young, < 1 Myr. Thus, $t_{\rm b,p}$ should be less than 1 Myr. Young clusters continue to undergo star formation for several million years until they either deplete their original molecular gas reservoirs or until gas is expelled by feedback processes, ultimately leading to the cessation of star formation. As revealed in Fig.\ref{lose}, HII-clumps have not significantly dispersed their natal molecular gas. Thus, they should be very young.
The ratio of PMS stars with disks to protostars (also called the Class II/Class I ratio) for the cluster cores in \citet{Gutermuth2009-184} requires a period of sustained star formation that lasts 2–3.5 Myr \citep{Megeath2022-134}. The ages of the Orion Nebula Cluster (ONC) stars may in actuality span $1-4\,$Myr \citep{DaRio2010-722,Reggiani2011-534,Getman2014-787-109,Beccari2017-A22, Jerabkova2019-A57}. Therefore, we should consider different values of $\sigma_{\rm t}$ to cover possible time spans of star formation.
Finally, we took $t_{\rm b,p}= 0.1, 0.25, 0.5, 1$ Myr and $\sigma_{\rm t}=$ 0.25, 0.5 and 1 Myr. There are a total of 12 cases here.

The typical values of the young protostellar stage is ~0.5 Myr ($\approx 0.1$ Myr for the Class 0 phase and $\approx 0.44$ Myr for the Class I phase) and that of the Class II phases is $\approx 2$ Myr \citep{Evans2009-181, Hsieh2013-205, Pokhrel2020-896, Megeath2022-134, Dib2024arXiv}.
In this work, only the protostars with ages less than 0.5 Myr are used to calculate the accretion luminosity.

In Fig.\ref{flat}, for the case of a flat SFH, nearly 90\% stars are excluded. Thus, 
the case with a flat SFH can not account for difference in luminosity. In the cases with time-dependent SFHs, 
as shown in Fig.\ref{g1} and Fig.\ref{g2}, as long as 20\% of the stars within the embedded cluster are still accreting, the contribution of accretion luminosity will be significant for low-mass clumps.
Currently, we have assigned corresponding isochrones based on the age of each star, rather than using the 0.1 Myr isochrone for all stars in Fig.\ref{fit}(b). As shown in Fig.\ref{g2}, the fitting lines remain comparable to Fig.\ref{fit}(b). For low-mass HII-clumps, the contribution of the stellar luminosity to the total luminosity is minimal. For high-mass HII-clumps, isochrones of different ages yield consistent results. Therefore, in the above analysis, the choice of isochrones is not important.

\subsection{The final star formation efficiency of the clump}\label{SFE}
\begin{figure}
\centering
\includegraphics[width=0.48\textwidth]{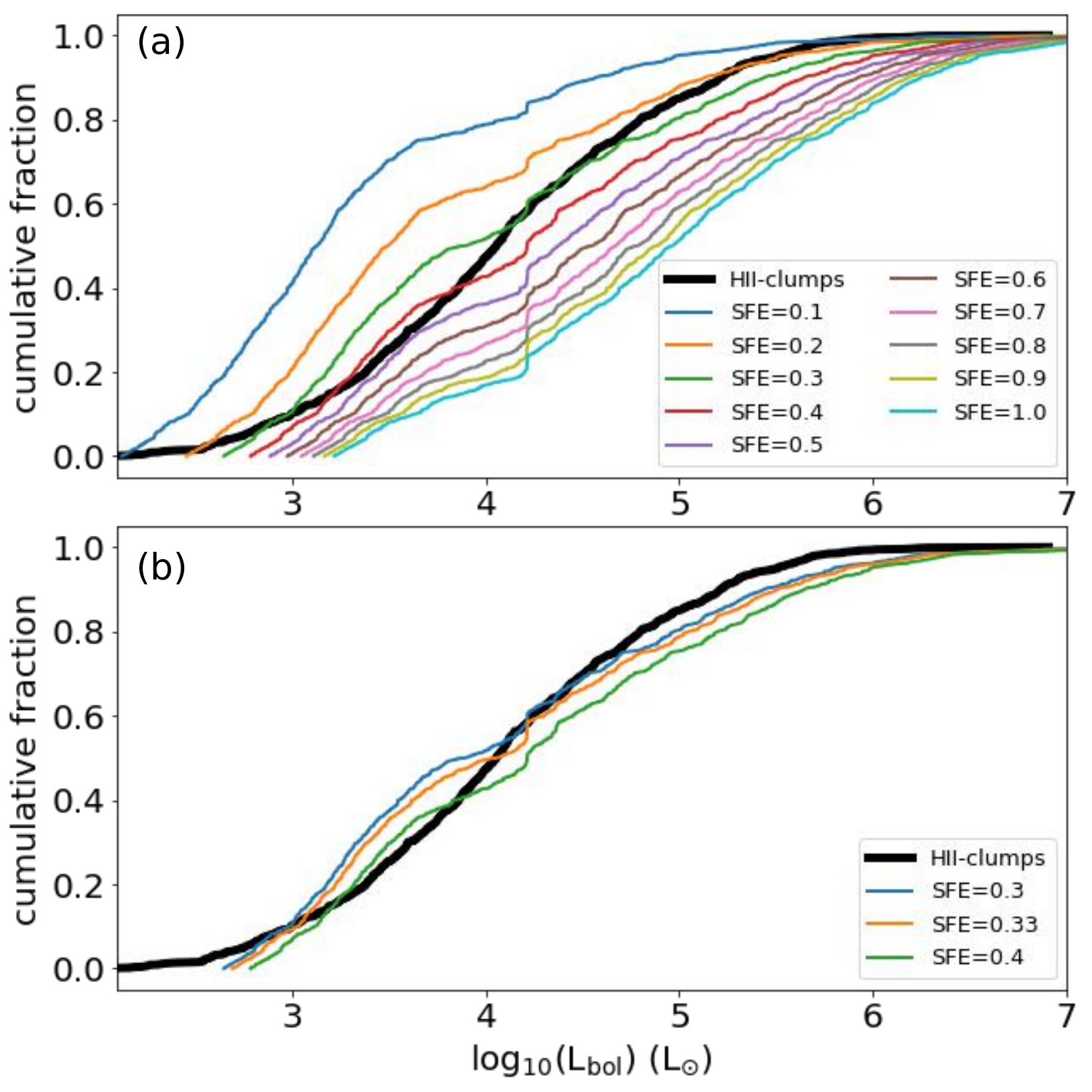}
\caption{
Cumulative distributions of log$_{10}$(L$_{\rm bol, obs}$) of HII-clumps (black thick lines) and log$_{10}$(L$_{\rm bol, ecl}$) of synthetic embedded clusters under different SFEs.}
\label{curve}
\end{figure}

\begin{table}
	\centering
	\caption{The median values of log$_{10}$(L$_{\rm bol, ecl}$) for the synthetic embedded clusters under different SFEs. "p" is the p-value of the Kolmogorov-Smirnov (KS) test between log$_{10}$(L$_{\rm bol, obs}$) and log$_{10}$(L$_{\rm bol, ecl}$).}
	\label{line}
	\begin{tabular}{cccc} 
		\hline
  SFE	&	p	&	median 	\\
  \hline
0.32	&	0.00003	&	4.004	\\
0.33	&	0.00009	&	4.056	\\
0.34	&	0.00038	&	4.103	\\
0.35	&	0.00164	&	4.145	\\
0.36	&	0.00195	&	4.178	\\
0.37	&	0.00055	&	4.202	\\
0.38	&	0.00038	&	4.212	\\
0.39	&	0.00007	&	4.213	\\
0.4	&	0.00003	&	4.215	\\
0.41	&	0.00002	&	4.218	\\
      \hline
	\end{tabular}
    \label{median}
\end{table}

In the above discussion, we adopted a constant SFE = 0.33 for all HII-clumps following \citet{Kroupa2001-321,Megeath2016-151,Yan2017-607,Banerjee2018-424,Yan2023-670}, although the masses of all HII-clumps can span more than 5 orders of magnitude.
In Sec.\ref{accretion}, by adding the accretion luminosity, the model is able to fit the observations. However, one could argue that the deviation between the model without the accretion luminosity and the observation in Fig.\ref{fit}(a) is due to assuming a constant SFE = 0.33. As described in \citet{Dib2013-436,Wells2022-516}, the SFE may change as a function of clump mass. 
If the SFE decreases with increasing mass of the clump, low-mass clumps with high SFE can also decrease the deviation between the model and the observed data in Fig.\ref{fit}(a). Thus, there is a critical question, for all HII-clumps with significantly different masses, whether they have similar final star formation efficiency? The good fit in Fig.\ref{fit}(b) supports the case of a SFE$\approx$0.33, but the scatter at any fixed cluster mass could be explained by a certain degree of variation of the SFE.

In order to evaluate the effect of the SFE for HII-clumps, we consider values of the SFE that fall in the range [0.1, 1] with a step of 0.01. Therefore, each HII-clump has 91 alternative SFEs, which correspond to 91 synthetic embedded clusters using the method in Sec.\ref{cluster}. In \citet{Yan2017-607}, the assumed lower limit of embedded cluster mass is M$_{\rm ecl,min}$ = 5 M$_{\odot}$. 
When SFE = 0.1, 924 HII-clumps in the sample in Sec.\ref{sample} can meet this requirement, i.e. M$_{\rm cl}$*0.1 > 5 M$_{\odot}$.
The median value of the observed bolometric luminosity log$_{10}$(L$_{\rm bol, obs}$) for 924 HII-clumps is 4.057. Table.\ref{median} displays the median values of the total bolometric luminosity log$_{10}$(L$_{\rm bol, ecl}$) for the synthetic embedded clusters under different SFEs.
When SFE = 0.36, the p-value of the Kolmogorov-Smirnov (KS) test between log$_{10}$(L$_{\rm bol, obs}$) and log$_{10}$(L$_{\rm bol, ecl}$) has the highest value, although this value is smaller than 0.05. The corresponding median value of log$_{10}$(L$_{\rm bol, ecl}$) is 4.178, comparable with that of log$_{10}$(L$_{\rm bol, obs}$). 
Interestingly, when SFE = 0.33, the median value of log$_{10}$(L$_{\rm bol, ecl}$) is almost the same with that of log$_{10}$(L$_{\rm bol, obs}$).
In Fig.\ref{curve}(a) and (b), when SFE$\approx$0.33, the cumulative distributions of log$_{10}$(L$_{\rm bol, obs}$) and log$_{10}$(L$_{\rm bol, ecl}$) are also comparable. 
In summary, SFE = 0.33 can be a typical value for all HII-clumps.

However, the analysis presented here follows the procedure in Sec.\ref{accretion}. We do not consider the star formation history and still assume all stars are in accretion. Thus, we should be cautious about the results. 
Low-mass HII-clumps with high SFE may lead to the excess luminosity shown in Fig.\ref{fit}. The current analysis can not exclude this possibility.

\section{Discussion and conclusions}

In Fig.4 of \citet{Hosokawa2009-691}, when M$_{*}$>10 M$_\odot$, L$_{*}$ begins to exceed L$_{\rm acc}$. Thus, for high-mass stars, the contribution from L$_{*}$ to L$_{\rm bol, str}$ is more important than L$_{\rm acc}$.
For the synthetic embedded clusters, when log$_{10}$(M$_{\rm cl}$/M$_{\odot}$)
$\approx$ 2.2, stars with masses larger than 10 M$_{\odot}$ begin to appear. In Fig.\ref{fit}(b), we can see the black line indeed begins to approach the red line around log$_{10}$(M$_{\rm cl}$/M$_{\odot}$)
$\approx$ 2.2. But now, the number of low-mass stars is still much larger than high-mass stars in the synthetic embedded cluster, the total accretion luminosity of low-mass stars remains dominant. When log$_{10}$(M$_{\rm cl}$/M$_{\odot}$)
$\approx$ 2.8, the luminosity of high-mass stars becomes dominant and L$_{\rm acc}$ can be ignored. Therefore, there is a mass limit of log$_{10}$(M$_{\rm cl}$/M$_{\odot}$)
$\approx$ 2.8, when the HII-clump mass is less than this limit, one must consider the contribution of the 
accretion luminosity from protostars to the observed bolometric luminosity L$_{\rm bol, obs}$ of the HII-clump.

In the above analysis, we employed classical pre-main sequence contraction tracks, but it needs to be kept in mind that the true path  an accreting pre-main sequence star takes in the Hertzsprung-Russell diagram (HRD) is likely to differ substantially as its internal structure, mass and current radius will reflect the accretion history \citep{Wuchterl2003-398}. The accretion rates onto protostars are time-variable \citep{Wuchterl2003-398,McKee2003-585,Schmeja2004-13,Dib2010-405,Fischer2017-840} and this may lead to a dispersion of the data points in Fig.\ref{fit}. Moreover, although our results seem to indicate that a constant SFE=0.33 is quite plausible, variations in the SFE at any given clump mass, along with potential variations of the IMF, could also help explain the scatter in the L$_{\rm bol, obs}$-M$_{\rm cl}$ relation.
Moreover, Low-mass HII-clumps with high SFE may also explain the excess luminosity. Compared to the very limited analysis in Sec.\ref{SFE}, a comprehensive discussion on the impact of SFE must take into account the star formation history. However, the star formation history of HII-clumps is quite uncertain, which will make the estimation of SFE quite uncertain as well. Furthermore, SFE is just one of many factors influencing the total luminosity of HII-clumps. We cannot simply explain the observational data by increasing SFE alone. Various factors must be considered together, which is certainly complex.

In this work, we demonstrate that the accretion process has important contribution to the total luminosity of low-mass HII-clumps. However, many HII-clumps exhibit significantly higher luminosity than the fitting lines in Fig.\ref{g2} and Fig.\ref{g3}, which indicates that the accretion should not be the only important factor. Other physical processes, such as the external heating from HII regions and the flaring from PMS stars and protostars \citep{Flaccomio2018-620,Zhang2020-637,Getman2021-916}, may also contribute to the total luminosity.

\begin{acknowledgements}
We would like to thank the referee for sharing detailed comments and suggestions that helped improve and clarify several aspects of this work.
\end{acknowledgements}


\bibliographystyle{aa} 
\bibliography{ref}


\appendix

\end{document}